\documentclass[showpacs,aps,prb,reprint,twocolumn]{revtex4-1}

\usepackage{graphicx}
\usepackage{dcolumn}
\usepackage{bm}
\usepackage{siunitx}
\usepackage{hyperref}
\usepackage{color}
\usepackage[dvipsnames]{xcolor}
\usepackage{float}
\usepackage{amsmath}

\hypersetup{
colorlinks=true,
urlcolor=blue,
citecolor=black,
linkcolor=black
}

\usepackage{commath}
\usepackage{amsmath,bm}
\usepackage{hyperref}
\usepackage{txfonts}

\draft 

\begin{document}

\title
{
Josephson junctions as nanoelectromechanical Bloch oscillators
}

\author{Thomas McDermott}

\affiliation
{
School of Physics and Astronomy, University of Exeter, EX4 4QL, Exeter, United Kingdom
}

\author{Hai-Yao Deng}
\thanks{H.-Y. Deng and T. McDermott contributed equally to this work.}
\affiliation{School of Physics and Astronomy, Cardiff University, 5 The Parade, Cardiff CF24 3AA, Wales, United Kingdom}

\author{Eros Mariani}

\affiliation
{
School of Physics and Astronomy, University of Exeter, EX4 4QL, Exeter, United Kingdom
}

\begin{abstract}
\noindent 

Bloch oscillations were predicted and observed long ago in Josephson junctions, which are thin insulating or conducting layers bridging two superconductors. Here we forward a scheme for utilizing these oscillations to amplify nano-metric mechanical vibrations. Highly coherent non-classical mechanical states can be produced and controlled by simply applying a DC bias current and an in-plane magnetic field. The vibrations induce peaks in the I-V characteristic of the junction and can be detected by simple DC voltage measurements.

\end{abstract}

\maketitle


When subjected to a constant and uniform electric field~\cite{zener}, electrons in solids may undergo periodic motions known as Bloch oscillations. These oscillations can be revealed either in the transport properties of the solid~\cite{leo,dekorsy} or the radiation generated by them~\cite{waschke}. One of the most practical aspects of Bloch oscillations is that they can be used to amplify radiation in the terahertz range~\cite{unterrainer}. 

Bloch oscillations are not exclusive to solids. In their pioneering work, Likharev \emph{et al.} showed~\cite{likharev,averin,likharev2,averin2} that the dynamics of a Josephson junction - a weak link made of non-superconducting material bridging two superconductors - was formally equivalent to that of an electron moving in a one-dimensional solid. As such, Bloch oscillations also occur in these junctions when driven by a DC current, which plays a similar role as the constant uniform electric field that induces ordinary electronic Bloch oscillations. Bloch oscillations in a Josephson junction can be revealed in its I-V characteristic in the presence of a probing AC current. However, they are not associated with any electromagnetic radiation and their direct observation has not been achieved~\cite{kuzmin,vora}.

In this theoretical work, we analyze a device in which a Josephson junction functions as a Bloch oscillator amplifying nano-scale mechanical vibrations in the junction, which resembles an electronic Bloch oscillator amplifying electromagnetic radiation placed in a single-mode optical cavity. In our device, the coupling between the Bloch oscillations and the mechanical vibrations is realized by an external in-plane magnetic field. The vibrations can be controlled by tuning the Bloch oscillation frequency via varying the DC current. They can be switched on or off by tuning the Bloch oscillation frequency to or away from the mechanical resonance frequency. This is reminiscent of the so-called Bloch gain~\cite{unterrainer}, whereby radiation can be absorbed or emitted by an electronic Bloch oscillator, depending on the offset between the oscillation frequency and the radiation frequency.

A significant aspect of our results is their potential use in generating highly coherent non-classical mechanical vibration states. Such states have attracted a lot of interest in recent years for allowing fundamental tests of quantum mechanics and their applications in mass, position and force measurements with unprecedented precision~\cite{schwab,bachtold}. As to be shown in what follows, the number of quantum of vibrations (called vibron) grows steeply from zero to over hundreds once the bias current exceeds a threshold value that depends on the applied magnetic field. Meanwhile, the spectral width of the vibrations obviously sharpens, indicating the onset of 'lasing' phenomenon in the system.

\begin{figure}
    \includegraphics[width=85mm]{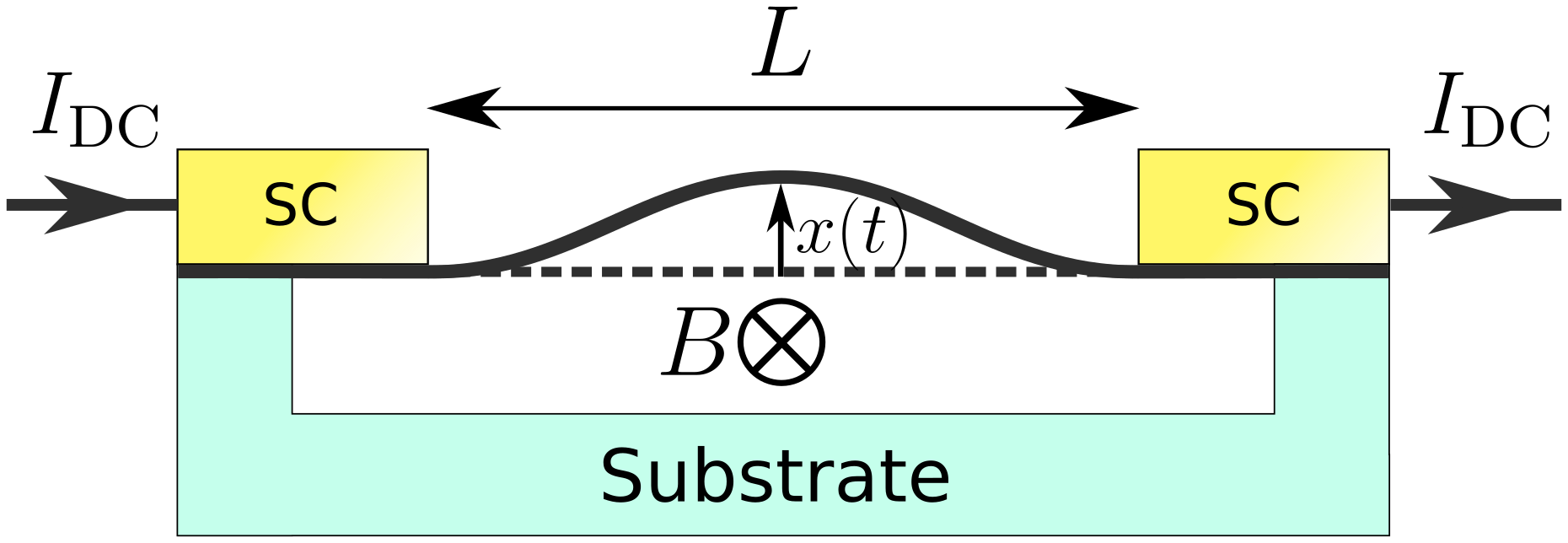}
    \caption{A Josephson junction formed by an electromechanical resonator suspended between two superconducting contacts. A bias current $I_\text{DC}$ through the device sets up Bloch oscillations which are coupled to mechanical vibrations through an in plane magnetic field $B$.}
    \label{setup}
\end{figure}

Our device is sketched in Fig.~\ref{setup}, where a mechanical resonator with oscillator mass $M$, effective length $L$ and fundamental resonance frequency $\omega_0$ is suspended between two superconducting electrodes. The device operates as a Josephson junction with a critical current $I_\text{c}$ and capacitance $C$. It is then biased by a DC current $I_\text{DC}$. An in-plane magnetic field $B$ can be applied to couple the electronic current in the device and the mechanical oscillations via the Lorentz force.

Quantum mechanically, a Josephson junction has two dynamical variables, the charge $\hat{Q}$ and the gauge invariant phase $\hat{\varphi}$, which obey the canonical commutation relation $[\hat{\varphi},\hat{Q}/2e]=i$, with $2e$ being the charge of a Cooper pair. In the uncoupled case, where $B = 0$, the dynamics of the junction is governed by the following Hamiltonian~\cite{likharev},
\small
\begin{equation}
    \hat{H}_0(\hat{Q},\hat{\varphi};C) = \frac{\hat{Q}^2}{2C} - \frac{\hbar I_c}{2e} \cos (\hat{\varphi})+ \frac{\hbar}{2e}(\hat{I_q} - I_\text{DC})\hat{\varphi} + \hat{H_q},
\label{Huncoupled}
\end{equation}
\normalsize
\noindent where $I_q$ is the normal current carried by thermally excited quasi-particles, which serve as a heat bath for the Cooper pair condensate, and $\hat{H}_q$ is the Hamiltonian describing these quasi-particles. As aforementioned, the Hamiltonian $\hat{H}_0$ also describes a quantum particle moving in a periodic potential, with the mass, position and momentum of the particle corresponding to $C$,  $\hat{\varphi}/2e$ and $\hat{Q}$, respectively. In addition, the impressed current $I_\text{DC}$ acts as an external electric field for the particle while $I_q$ acts as a thermally induced field. 


Switching on the magnetic field, a Lorentz force due to the electronic current in the device results in a coupling to the mechanical vibrations. As shown in our previous work~\cite{mcdermott}, this coupling affects the junction in two ways. Firstly, it renormalizes the capacitance to $\tilde{C} = C/(1 + \mu^2)$ and the mass to $\tilde{M} = M(1 + \mu^2)$, where $\mu = B/B_0$ with $B_0 = \sqrt{MC^{-1}}/L$. Secondly, it gives rise to a gauge field proportional to the mechanical momentum $\hat{p}$. The total Hamiltonian of the system now reads
\begin{equation}
    H = \hat{H}_0(\hat{\pi},\hat{\varphi};\tilde{C}) + \frac{\hat{p}^2}{2\tilde{M}} + \frac{1}{2}\tilde{M}\tilde{\omega}_0^2 \hat{x}^2 + \hat{H_b} + \hat{H_i},
\label{Hcoupled}
\end{equation}
\noindent where $\hat{\pi} = \hat{Q} - \frac{\tilde{C}BL}{M}\hat{p}$, $\tilde{\omega}_0 = \omega_0/(1+\mu^2)$, $\hat{x}$ is the displacement of the resonator relative to the constant slack $B I_\text{DC}L/M\omega_0^2$, $H_b =  \sum_\omega \hbar \omega (\hat{b}_\omega^\dagger \hat{b}_\omega + 1/2)$ models a mechanical heat bath consisting of a spectrum of harmonic oscillators and $\hat{H_i} = \sum_\omega \hbar \lambda_\omega (\hat{a} \hat{b}_\omega^\dagger + \hat{a}^\dagger \hat{b}_\omega)$ describes the coupling to this bath. Here $\hat{b}_\omega$, $\hat{b}_\omega^\dagger$ are the annihilation and creation operators for the heat bath oscillators, while $\hat{a}$ and $\hat{a}^\dagger$ are those for the resonator, yielding $\hat{x} = \sqrt{\frac{\hbar}{2M\omega_0}}(\hat{a}+\hat{a}^\dagger)$ and $\hat{p} = i \sqrt{\frac{\hbar M \omega_0}{2}} (\hat{a}^\dagger-\hat{a})$. The vibron number operator is given by $\hat{n} = a^\dagger a$. The parameters $\lambda_\omega$ are chosen by assuming a white noise distribution for the heat bath, i.e.  $\sum_\omega \lambda_\omega^2 e^{i \omega (t-t')} = \Gamma \delta(t-t')$, so that the interaction with the heat bath leads to the usual velocity dependent damping.

Mathematically, the Hamiltonian (\ref{Hcoupled}) resembles that of an electronic Bloch oscillator placed in a single mode optical cavity with frequency $\omega_0$, with the current $\hat{I_q} - I_\text{DC}$ acting as an electronic scalar potential, while the mechanical resonator introduces an effective vector potential $\hat{p}$. The coupling strength $\tilde{C}BL/M$ is tunable by the field $B$.

The eigenvalues of the first two terms of the bare Hamiltonian (\ref{Huncoupled}) form Bloch bands with dispersion $E_s(q)$, where $s = 0, 1, ...$ is the band index and $q$ is called the quasi-charge playing the same role as the quasi-momentum in a real crystal lattice. $E_s(q)$ is a periodic function of $q$ with period $2e$. At low temperatures and for small $I_\text{DC}$, i.e. $k_B T \ll E_g$ and $\hbar I_\text{DC}/2e \ll E_g$, where $E_g = \text{min}(E_1 - E_0)$ is the band gap separating the lowest two bands, thermal excitations and Zener tunneling are suppressed so that we may restrict ourselves to the lowest energy band $s = 0$. As such, we can perform a semi-classical analysis of the system. By the standard procedures, the semi-classical equations of motion can be written down as follows
\begin{equation}
    \frac{\hbar \dot{\varphi}}{2e} = \frac{\partial E}{\partial q} = \frac{Q}{\tilde{C}} - \frac{BLp}{M}, \quad
    \dot{q} = I_\text{DC} - I_q - \frac{\tilde{C}BL}{M}\dot{p}.
    \label{phidot}
\end{equation}
\noindent Here the un-hatted symbols are the quantum-mechanical state averages of the corresponding operators, the dot over a quantity denotes a time derivative and $E(q) = E_0(q)$ is the dispersion of the band $s = 0$. The normal current $I_q$ carried by quasi-particles should be determined self-consistently. If the superconducting gaps of the electrodes are much larger than $I_\text{DC}/2e$, however, it takes on a simple form \cite{likharev}
\begin{equation}
    I_q = \frac{1}{R}\frac{\partial E}{\partial q},
    \label{iq}
\end{equation}
\noindent where $R$ is the normal electrical resistance of the junction, in accordance with the RCSJ model. Equations (\ref{phidot}) and (\ref{iq}) can readily be solved for $B = 0$ and have been extensively studied previously. As shown
by Likharev et al. \cite{likharev,averin,likharev2,averin2}, below a threshold current $I_t = \text{max}(\partial_q E/R)$ there exists a static Ohmic solution $\dot{q} = 0$, yielding the measured voltage by $V = Q/C = I_\text{DC}R$. Above $I_t$, however, q evolves with time, producing Bloch oscillations in the voltage with frequency
\begin{equation}
    \omega_B = \frac{\pi}{e}\left(I_\text{DC} - \frac{\langle V\rangle}{R}\right),
    \label{frequency}
\end{equation}
\noindent where the angle brackets indicate the time average. The oscillatory regime is characterized by a negative differential resistance $d\langle V\rangle/d I_\text{DC} < 0$. In the presence of an additional AC current bias, the I-V characteristic shows Ohmic branches where the Bloch oscillation frequency $\omega_B$ couples to the harmonics or sub-harmonics of the AC current frequency.

The equations of motion for the mechanical part are straightforward to obtain and are given by
\begin{equation}
    \dot{x} = \frac{p - CBL \partial_q E}{\tilde{M}}, \quad
    \dot{p} = -M\omega_0^2 x - 2M\Gamma\dot{x}.
    \label{xdot}
\end{equation}
\noindent Equations (\ref{phidot}) - (\ref{xdot}) are now closed and can be solved for $Q$, $\varphi$, $x$ and $p$.


The mechanical vibration amplitude turns out to be of the order of zero-point motion amplitude, $x_\text{zp} = \sqrt{\hbar/2M\omega_0}$. It is thus desirable to characterize the resonator also by the expectation value of the vibron number, $n(t)$. By Heisenberg's equation, one can show that  
\begin{equation}
\dot{n} = - \frac{BL\omega_0}{\hbar} \mathcal{C}_1 - 2\Gamma n, \label{n1}
\end{equation}
where $\mathcal{C}_1$ is the state average of $\hat{x}\hat{Q}$. Applying Heisenberg's equation to $\mathcal{C}_1$ and any new quantity that appears in the ensuing equations, one finds
\begin{eqnarray}
\dot{\mathcal{C}}_1 &=& x (I_\text{DC} - I_q) + \mathcal{C}_2/M - BL\mathcal{C}_3/M - \Gamma \mathcal{C}_1 , \\
\dot{\mathcal{C}}_2 &=& p(I_\text{DC}-I_q) - M\omega^2_0 \mathcal{C}_1 - \Gamma \mathcal{C}_2 , \\
\dot{\mathcal{C}}_3 &=& 2Q(I_\text{DC}-I_q) , \label{n4}
\end{eqnarray}
where $\mathcal{C}_2$ and $\mathcal{C}_3$ are the state averages of $\hat{p}\hat{Q}$ and $\hat{Q}^2$, respectively. In establishing these equations, we have used the semi-classical relation $d\hat{Q}/dt = I_\text{DC} - I_q$. Equations (\ref{n1}) - (\ref{n4}) can be solved for $n(t)$, with $x$, $p$ and $Q$ obtained from (\ref{phidot}) - (\ref{xdot}).

The band dispersion $E(q)$ has been well known. In the limit where the Josephson energy $E_J = \hbar I_c/2e$ is larger than the charging energy $E_C = e^2/2C$, one can employ a tight-binding approximation and find
\begin{equation}
E(q) = \frac{\Delta}{2} \left(1 - \cos(\pi q/e)\right),
\end{equation}
\noindent where $\Delta$ is the band width given by
\begin{equation}
\frac{\Delta}{E_J} = 16\sqrt{\frac{2}{\pi}}\exp\left(-\sqrt{\frac{8E_J}{E_C(1+\mu^2)}}\right).
\label{bandwidth}
\end{equation}
\noindent We numerically solve Eqs. (\ref{phidot}) - (\ref{n4}) using the following parameters appropriate for carbon nanotube devices $I_c = 1$nA, $R = 10$k$\Omega$, $C = 5\times10^{-14}$F, $L=1\mu$m, $M=10^{-21}$kg, $\omega_0/2\pi = 1$GHz, $Q_\Gamma = \omega_0/\Gamma = 10^4$, for which $E_J/E_C=1.29$ \cite{cleuziou,herrero,peng,garcia,huttel,laird,moser}.

\begin{figure}
    \includegraphics[width=85mm]{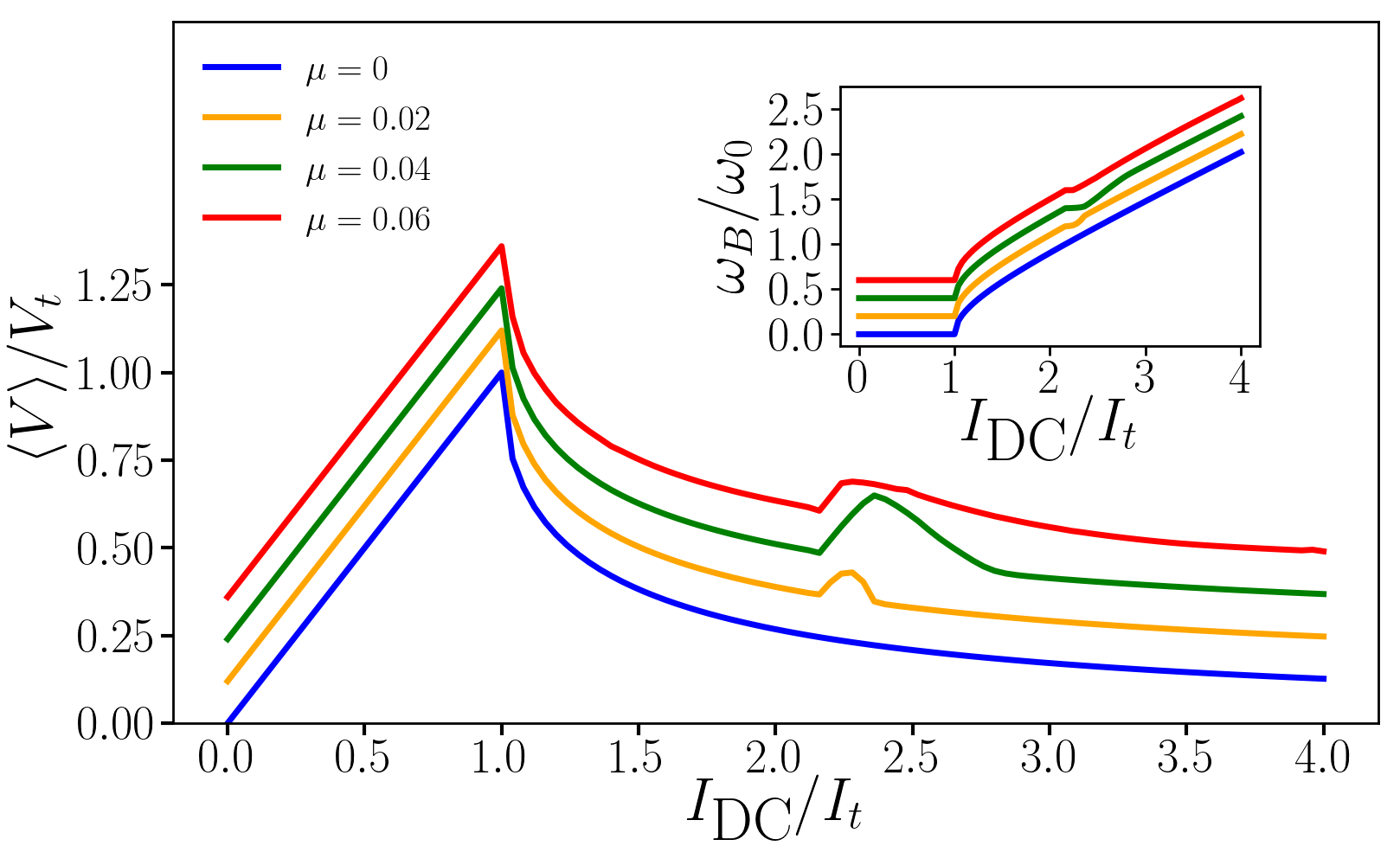}
    \caption{I-V characteristic obtained from Eqs. (\ref{phidot}) - (\ref{xdot}) for various values of coupling strengths $\mu = B/B_0$ with the parameters quoted in the main text, for which $I_t = 0.167 \si{\nano \ampere}$, $V_t = I_t R = 1.67 \si{\micro \volt}$, $\omega_0/2\pi = 1\si{\giga \hertz}$, $B_0 = 141\si{\tesla}$. A distinct voltage peak of size $\sim 0.2\si{\micro\volt}$ is observable when the Bloch oscillation frequency $\omega_B$ locks onto the mechanical resonance frequency $\omega_0$. The inset explicitly shows this frequency locking. Plots with $\mu \neq 0$ are displaced upward along the vertical axis slightly for clarity.}
    \label{varray}
\end{figure}

Fig.~\ref{varray} shows the predicted I-V characteristic for a number of different values of $\mu$. Here $\langle V\rangle$ is the DC component of the voltage and $V_t = I_t R$. In the absence of Josephson-mechanical coupling, i.e. $\mu = 0$, the usual Ohmic behaviour for $I < I_t$ is observed along with negative differential resistance for $I > I_t$. Switching on the coupling causes the oscillator to vibrate, resulting in distinctive voltage peaks in the I-V curve. These peaks develop when the Bloch oscillation frequency $\omega_B$ matches the resonance frequency $\omega_0$, due to the gauge field $\tilde{C}BL\dot{p}/M$ acting as an effective AC current. The inset of Fig.~\ref{varray} shows that the frequency $\omega_B$ is locked to $\omega_0$ over the resonant region,  similar to the familiar Shapiro plateau \cite{mcdermott,shapiro}. 

\begin{figure}
    \includegraphics[width=85mm]{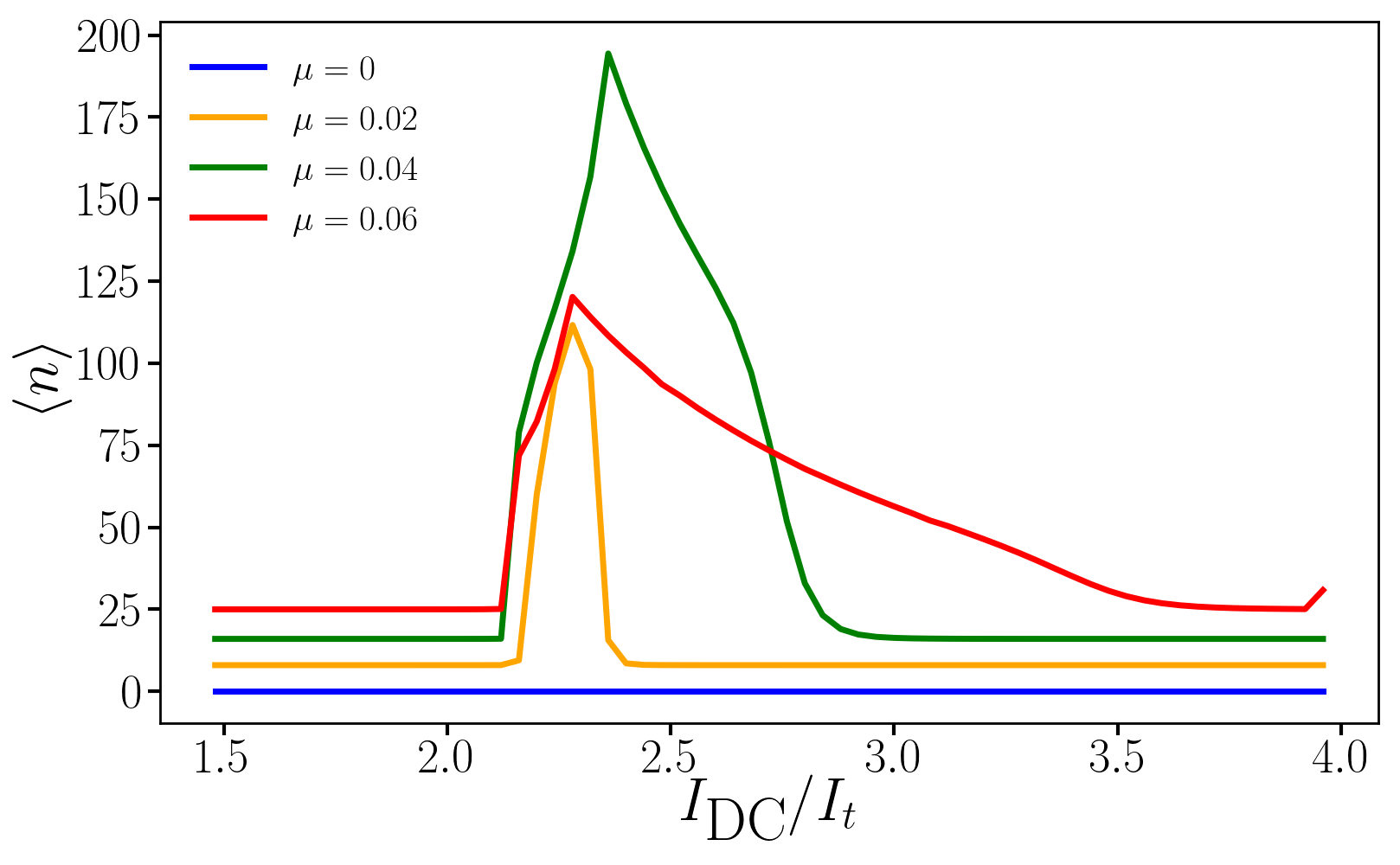}
    \caption{Time average of the vibron number $\langle n\rangle$ as a function of $I_\text{DC}$, obtained from Eqs. (\ref{n1}) - (\ref{n4}) for various values of coupling $\mu = B/B_0$ using the same parameters as in Fig.~\ref{varray}. Plots with $\mu \neq 0$ are displaced upward along the vertical axis for clarity.}
    \label{narray}
\end{figure}

Excitation of the oscillator can be further studied by looking at the evolution of the expectation value $n(t)$ of the vibron number operator $\hat{n}$. Solving Eqs. (\ref{n1}) - (\ref{n4}), we find that the oscillator initially set in the zero-vibron state eventually evolves into a state with $n(t)$ generally oscillatory but highly localized around an average $\langle n \rangle$. In Fig. \ref{narray} we display $\langle n\rangle$ as a function of the impressed current $I_\text{DC}$. A direct comparison with Fig.~\ref{varray} shows that the oscillator excitation coincides with the voltage peaks in the I-V characteristic. As such, the mechanical vibrations are directly manifested in the I-V curve. They can be turned on or off by simply tuning $\omega_B$ toward or away from $\omega_0$. Note that the oscillator is not excited gradually: increasing $I_\text{DC}$ does not increase $\langle n\rangle$ until frequency locking happens, thence the oscillator is suddenly activated with orders of magnitude increase of $\langle n\rangle$ from zero to the peak value $\sim 200$ for the parameters used here. This strong amplification of mechanical vibrations is similar to the case with a laser: in the latter an order of magnitude increase in the number of photons occurs once the lasing action sets in. The resulting mechanical state is thus highly coherent.

\begin{figure}
    \includegraphics[width=85mm]{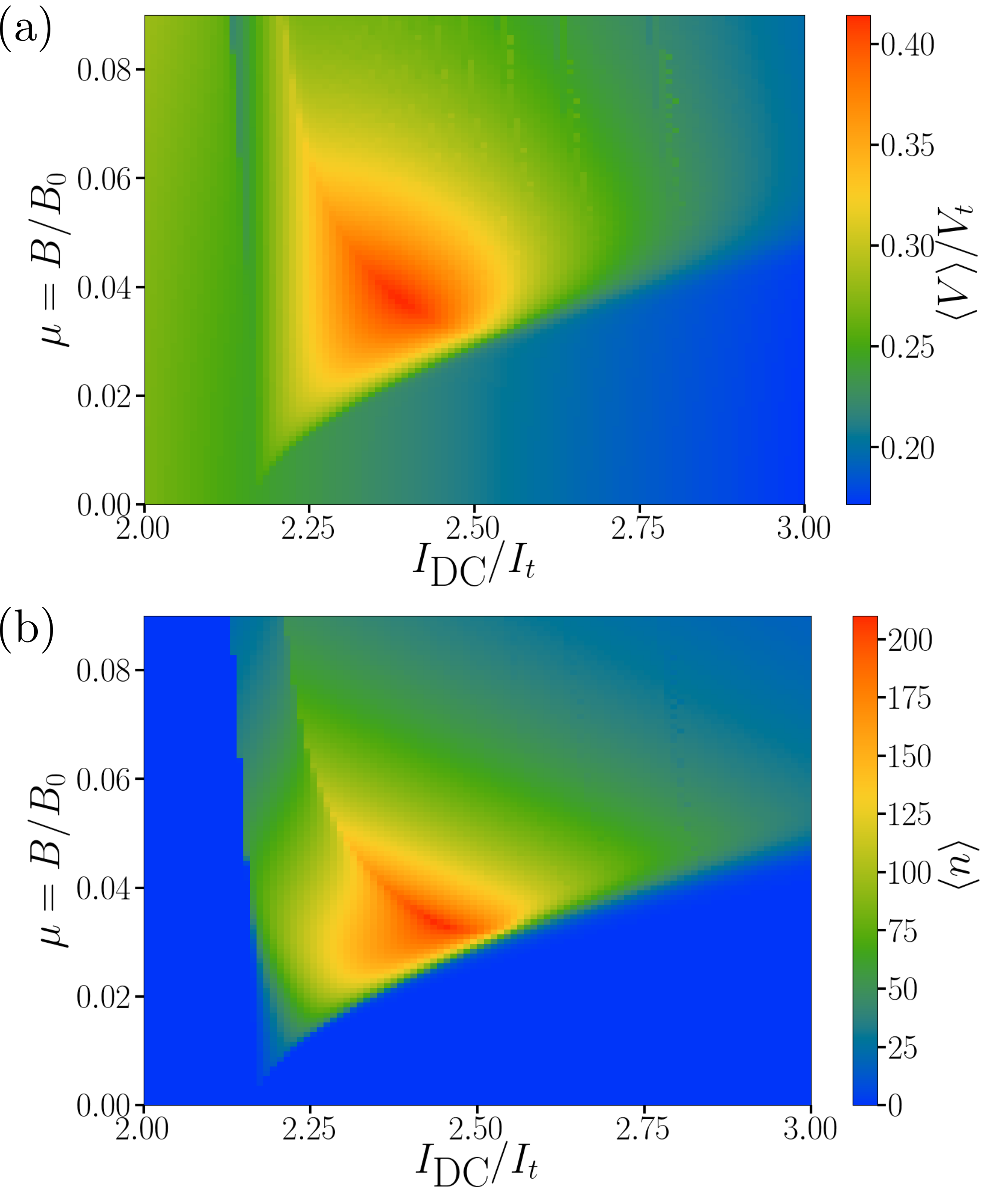}
    \caption{(a) DC Voltage $\langle V\rangle$ and (b) average of vibron number $\langle n\rangle$ versus $(I_\text{DC}/I_t,B/B_0)$ for the same parameters as in Figs. \ref{varray} and \ref{narray}. $\langle V\rangle$ and $\langle n\rangle$ peak about $1\mu$V and $200$, respectively.}
    \label{vgridngrid}
\end{figure}

To further establish the voltage hikes as fingerprints for the non-classical vibronic states, we show in Fig.~\ref{vgridngrid} the colour plots of $\langle V\rangle$ and $\langle n\rangle$ as functions of both the current and the magnetic field. Only the region of negative differential resistance tail, i.e. $I_\text{DC} > I_t$ is shown for clarity. A direct correspondence between $\langle V\rangle$ and $\langle n\rangle$ is observed. One sees that increasing the coupling $\mu$ or the impressed current $I_\text{DC}$ does not monotonically enhance the vibrations and increase $\langle n\rangle$. An optimal value of $(\mu,I_\text{DC})$ exists where $\langle n\rangle$ and $\langle V\rangle$ are peaked, see the red spot in the plots. Increasing these parameters beyond their optimal values begins to suppress the vibrations. The existence of an optimal $I_\text{DC}$ stems from the frequency locking discussed above. The existence of an optimal $\mu$ is likely due to the renormalization of the oscillator mass, which leads to smaller resonance frequency. Indeed, the red spot disperses to smaller optimal $I_\text{DC}$ and hence smaller $\omega_B$ as $\mu$ increases. 

Near the end, we make some remarks regarding the experimental realization of the scheme. Firstly, we see from the maps in Fig.~\ref{vgridngrid} that the peak values of $\langle V\rangle$ and $\langle n\rangle$ can be reached at $B\sim 4$T for the parameters used. This value of $B$ can be lowered by increasing $Q_\Gamma$ -- the mechanical quality factor, leading in the meanwhile to exalted peak values. Secondly, as the coupling strength $CBL/M$ is proportional to the capacitance $C$, it is desirable to increase $C$ to enhance the coupling. On the other hand, the peak value of $\langle V\rangle$ can be shown proportional to the band width $\Delta$. The latter should then be made bigger to get a significant voltage signal. As Eq. (\ref{bandwidth}) suggests, this would prefer a smaller $C$. For the best experimental outcome, $C$ should therefore be tuned somewhere between these two extremes. Finally, low temperatures are required for the measurements so as to ensure that the amplified mechanical vibrations exceed thermal fluctuations and the voltage peak does not get too much broadened. For the parameters given in the present work, we estimate that the temperature should be below $1\si{\kelvin}$. 

To conclude, we have shown that Bloch oscillations in Josephson junctions can be used to amplify mechanical vibrations and obtain coherent vibronic states of nano-scale resonators with only DC techniques. Our result not only provides a direct means of detecting Bloch oscillations in Josephson junctions and therefore complements the current understanding of Bloch oscillations, but may also find applications in for example precision measurements of mass, force and length.  


\section*{Acknowledgments}

We thank Saverio Russo for stimulating discussions. Financial support from the Leverhulme Trust (Research Project Grant RPG-2015-101), and the Royal Society (International Exchange Grant Nr. IE140367, Newton Mobility Grants 2016/R1 UK-Brazil, and Theo Murphy Award TM160190) are gratefully acknowledged.


\end{document}